\documentclass[aip,apl,%
  reprint,
  superscriptaddress]{revtex4-1}

\usepackage{graphicx}
\usepackage{epstopdf}
\usepackage{upgreek}
\usepackage{ifthen}
\usepackage[usenames,dvipsnames,svgnames,table]{xcolor}
\usepackage[abs]{overpic}

\newboolean{ShowComments}\setboolean{ShowComments}{true}  

\newcommand{\ShowMyCommnt}[1]{
\provideboolean{ShowComments}
\ifShowComments
  {#1} 
\fi
}
\newcommand{\NewComment}[2]{%
\expandafter\newcommand\csname #1Comment\endcsname[1]{\ShowMyCommnt{\textcolor{#2}{\textit{##1}}}}}

\NewComment{Harald}{blue}
\NewComment{Renate}{magenta}
\NewComment{Tilman}{green}

\newcommand{\einh}[1]{\,\mathrm{#1}}
\newcommand{\Figureref}[1]{Fig.~\ref{#1}}

\begin{document}

\title{Triple stack glass-to-glass anodic bonding for optogalvanic spectroscopy cells with electrical feedthroughs}

\author{R. Daschner}
\affiliation{5. Physikalisches Institut, Universit\"at Stuttgart, Pfaffenwaldring 57, 70550 Stuttgart, Germany}
\author{H. K\"ubler}
\affiliation{5. Physikalisches Institut, Universit\"at Stuttgart, Pfaffenwaldring 57, 70550 Stuttgart, Germany}
\author{R. L\"ow}
\affiliation{5. Physikalisches Institut, Universit\"at Stuttgart, Pfaffenwaldring 57, 70550 Stuttgart, Germany}
\author{H. Baur}
\affiliation{Institut f\"ur Gro\ss fl\"achige Mikroelektronik, Universit\"at Stuttgart, 70569 Stuttgart, Germany}
\author{N. Fr\"uhauf}
\affiliation{Institut f\"ur Gro\ss fl\"achige Mikroelektronik, Universit\"at Stuttgart, 70569 Stuttgart, Germany}
\author{T. Pfau}
\email{t.pfau@physik.uni-stuttgart.de}
\affiliation{5. Physikalisches Institut, Universit\"at Stuttgart, Pfaffenwaldring 57, 70550 Stuttgart, Germany}

\begin{abstract}
We demonstrate the use of an anodic bonding technique for building a vacuum tight vapor cell for the use of Rydberg spectroscopy of alkali atoms with thin film electrodes on the inside of the cell. The cell is fabricated by simultaneous triple stack glass-to-glass anodic bonding at $300\einh{^\circ C}$. This glue-free, low temperature sealing technique provides the opportunity to include thin film electric feedthroughs. The pressure broadening is only limited by the vapor pressure of rubidium and the lifetime is at least four months with operating temperatures up to $230\einh{^\circ C}$.
\end{abstract}

\maketitle

\noindent
Atomic vapor cells at and above room temperature are the basis for applications such as optical frequency references \cite{DAVLL}, magnetic \cite{polzik2010} and electric \cite{sed2012} field sensors,  and atom clocks \cite{kna2005}. Due to advances in miniaturization and fabrication techniques commercial products are now available \cite{csac}.
With the demonstration of Rabi flopping at GHz frequencies in a thermal vapor \cite{hub2011} and the possibility to excite Rydberg atoms in tight confinements \cite{kue2010} it has been shown that applications in nonlinear and quantum optics are possible in thermal vapor cells.
The large Stark effect associated with the huge polarizability of Rydberg atoms can be exploited to switch and modulate the optical properties of such cells \cite{mohapatra2008}.
In fact electric field control is also important to avoid drifts, broadening, and dephasing of the spectroscopy lines. While thin film based electrical feedthroughs are compatible with microfabrication techniques and can easily be scaled up to large numbers, they are destroyed when vapor cells are sealed above the melting point of glass.
Our previously demonstrated gluing technique \cite{das2012} is compatible with these requirements, but the maximum operating temperature was limited to $100\einh{^\circ C}$. This is a severe drawback when it comes to highly integrated cells and therefore to small vapor volumes. To achieve reasonable optical density in such thin cells, higher temperatures are required. Therefore it is necessary to develop techniques to produce high vacuum tight vapor cells with conductive, thin film based feedthroughs, that can withstand temperatures up to $200\einh{^\circ C}$ and under these conditions provide lifetimes that are long enough for commercial applications.

\begin{figure}[t]
\centerline{\includegraphics[width=8.4cm]{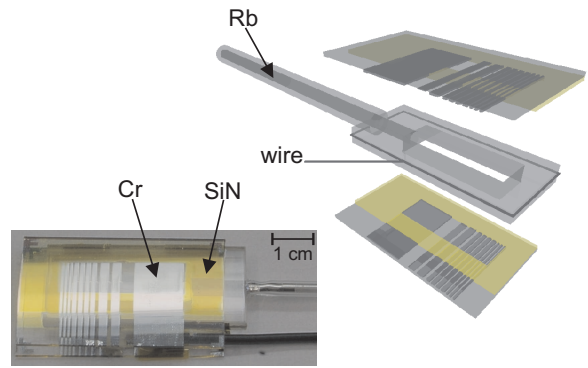}}
\caption{The cell consists of a glass frame with an attached tube, which acts as a reservoir for the rubidium (Rb). Two glass plates coated with chromium (Cr) and silicon nitride (SiN) are anodically bonded to the frame to enclose a cell volume of $1.5\einh{cm^3}$. Conductive silver is pasted around the frame and a wire is wrapped on top of the silver, to apply a high voltage in the middle of the frame to simultaneously bond the upper and the lower plate. 
\label{cell_picture}}
\end{figure}

Most of the currently available techniques to build high vacuum tight vapor cells are often not compatible with thin film coatings. The conventional sealing of glass cells by flame can not be used, as thin film electrodes would not survive the high temperatures required to melt glass. Other techniques like direct bonding also require high temperatures and polished surfaces with a flatness better than the height of the electrodes.
For glass frit bonding a technology to print and cure the glass frit is needed and only with a getter material inside a low background pressure can be achieved \cite{spa2005}.
Gluing with an epoxy has already been tried successfully \cite{das2012}, but even glues with a low outgassing rate lead to a pressure broadening over time, especially at high temperatures.
Due to this outgassing the cell could only be used for a few weeks and for temperatures below $100\einh{^\circ C}$.
Our new approach is an anodic bonding technique \cite{wal1969} which is similar to the one used to build vapor cells for chip scale atomic clocks \cite{kna2005}.  Instead of silicon-to-glass bonding we are using glass-to-glass bonding to attach two glass plates to a center glass frame. In this way we can easily fuse a glass tube to the frame to fill the cell with rubidium. In the case of glass-to-glass bonding, a thin film layer of silicon nitride is needed at the interface to form a permanent bond between the silicon of the thin film and the oxide of the glass \cite{ber2000}. Simultaneous bonding of both sides is necessary to avoid debonding of the first side while bonding the second one \cite{tin2010}. With this technique we can include metal feedthroughs below the bonding layer and still get low enough surface roughness to result in a perfect bond as long as the metal layers are thin enough.

The cell consists of a frame and two structured glass plates, see \Figureref{cell_picture}. The 
glass plates are coated with $20\einh{nm}$ of chromium which is photolithographically structured to small stripes. They can be contacted from the outside and used as electrodes to apply electric fields inside the cell. On top of the chromium layer we sputtered a layer of $200\einh{nm}$ silicon nitride in the shape of the glass frame. 
The $5\einh{mm}$ thick glass frame has a hole drilled from the side where a glass tube is fused to. This tube will later be used to fill the cell with rubidium. The top and the bottom surface of the frame are polished for $1.5\einh{h}$ each to remove deformations due to the melting.

\begin{figure}[b]
(a)
\centerline{\includegraphics[width=8.4cm]{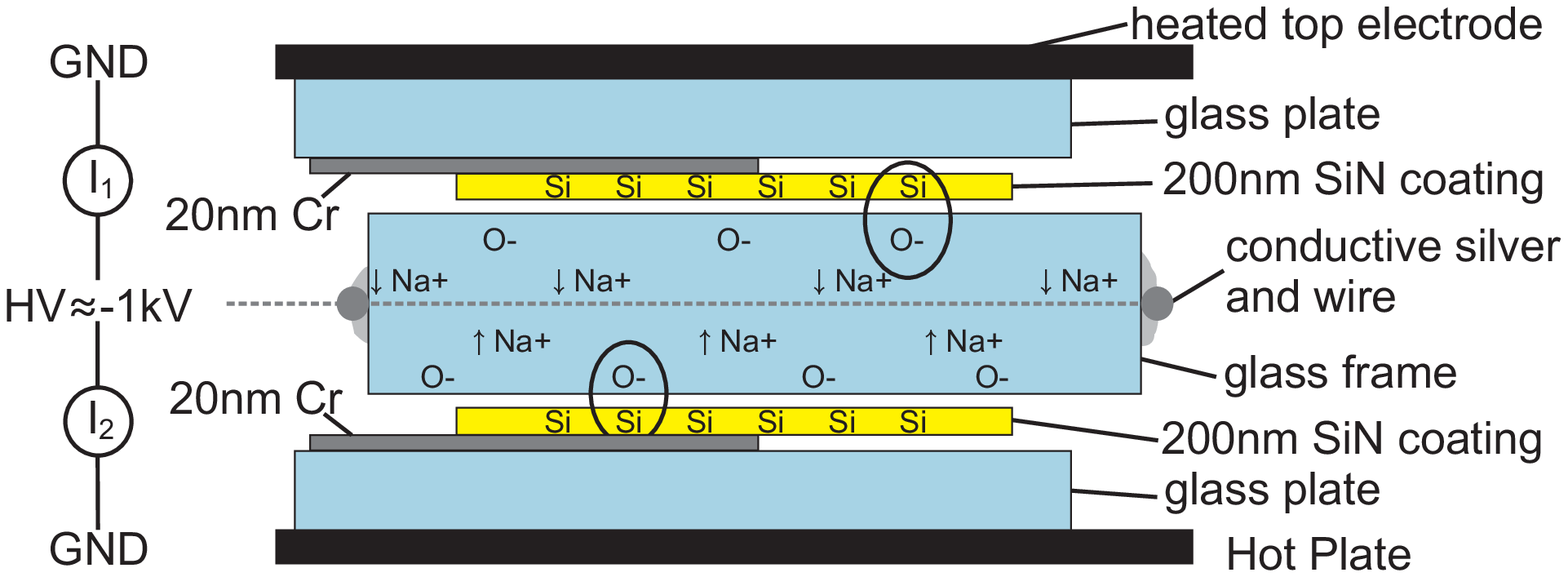}}\\
(b)
\centerline{\includegraphics[width=8.4cm]{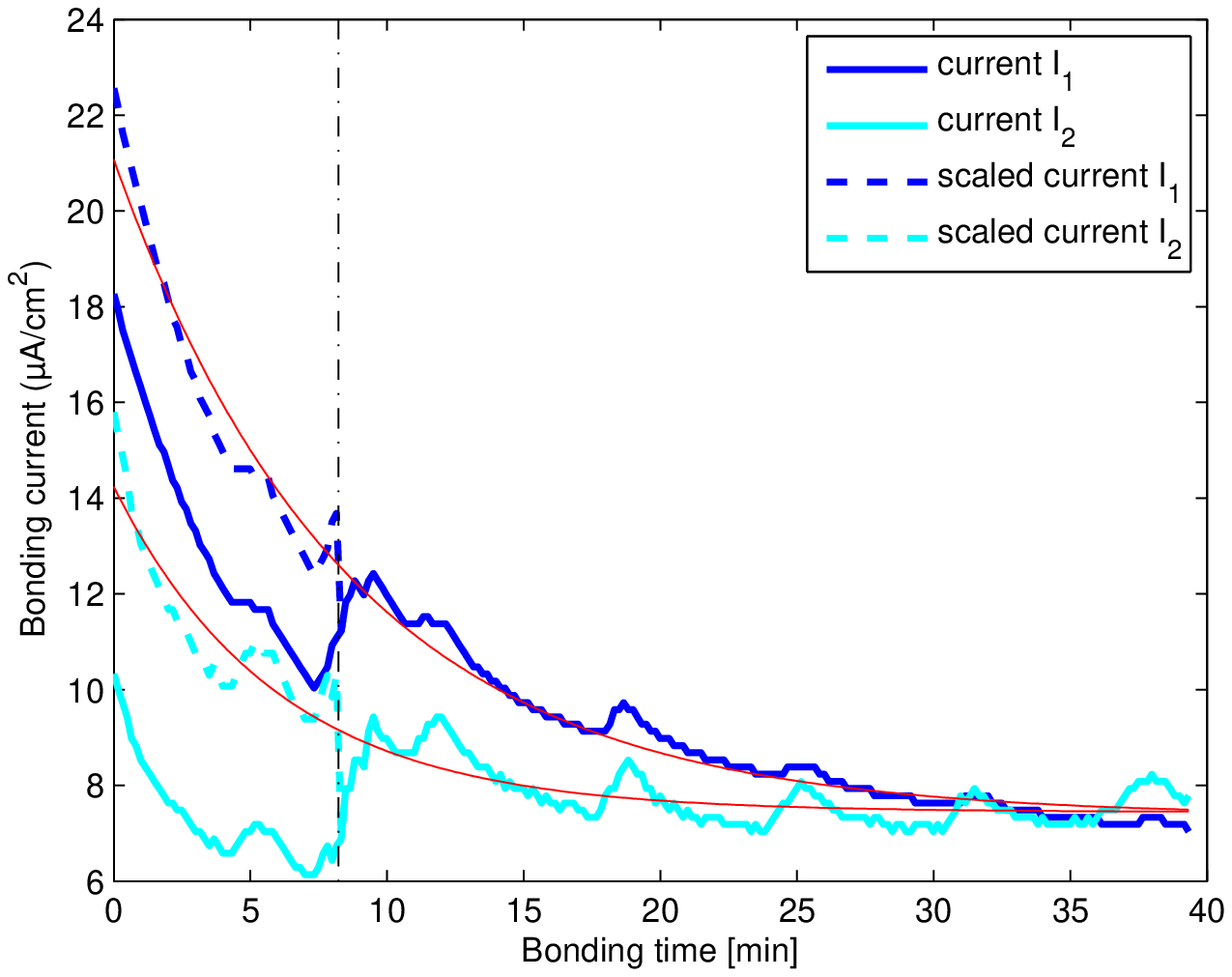}}\\
\caption{(a) Ion processes during anodic bonding. Sodium ions from the glass move towards the wire and leave behind open oxygen bonds. The moving charges build up a static electric field which presses the plates to the frame. The oxygen atoms can bond to the silicon atoms from the coating to form a permanent bond. (b) Current flowing through the glass during bonding on both sides. The first $8\einh{min}$ are bonded at $-850\einh{V}$, then the voltage is increased to $-990\einh{V}$. The first $8\einh{min}$ are scaled as described in the main text. The time constants are $8.6\einh{min}$ ($I_1$) and $6.0\einh{min}$ ($I_2$). 
\label{current}}
\end{figure}

For the anodic bonding, conductive silver and a wire are placed around the glass frame shown in \Figureref{cell_picture}. The glass plates are cleaned carefully with acetone and alcohol to remove all dust between the plates. The three glass parts are stacked with a steel electrode and an additional heater on top, as shown in \Figureref{current}(a).  On the one hand, it is highly favorable to reduce the bonding temperature as far as possible as a lot of coatings cannot withstand common bonding temperatures of $400\einh{^\circ C}$ to $800\einh{^\circ C}$. On the other hand, lower temperatures reduce the conductivity of the glass therefore the bonding time is increased and a reduced bond strength is observed.
The whole stack is homogeneously heated to the bonding temperature of only $300\einh{^\circ C}$.
When the complete stack is in thermal equilibrium a high voltage of $-850\einh{V}$ to $-990\einh{V}$ is applied at the wire on the middle of the glass frame.The hot plate and the upper electrode are held at ground.
Even at this low temperature ions can diffuse through the glass. Due to the polarity of the high voltage, the positively charged sodium ions from the glass frame move from both sides towards the wire in the middle of the frame, leaving open oxygen bonds at both interfaces between the plates. The sodium ions from the top and bottom glass plates also move towards the frame but the silicon nitride coatings act as diffusion barriers for these ions. The positively charged sodium ions build up on one side while the remaining negatively charged oxygen build  up on either side of the interface. These charges are forming an electric field, which presses the plates from both sides to the frame \cite{kno2006}. The open oxygen bonds from the frame can now form a molecular bond with the neighboring silicon atoms from the silicon nitride coating. This leads to a permanent, robust, and high vacuum tight bond between the two glass plates and the frame.

The current from the wire through the stack is measured during the bonding sequence on both sides. The first $8\einh{min}$ were done at $-850\einh{V}$. Then the voltage was increased to $-990\einh{V}$ to compensate for the fast decay of the current (see \Figureref{current}(b)). 
The amplitude of the current is different for the two sides. This can easily happen if there is a small temperature difference or when the surface flatness on both sides is slightly different. The currents should be kept at the same order of magnitude when simultaneously bonding both sides to avoid a different bond strength on the two sides. Only then are both sides bonded without defects.
The large total glass thickness of $7\einh{mm}$ used here as compared to the $0.5\einh{mm}$ glass plates used in common anodic bonding processes \cite{kno2006}, increases the bonding time from tens of seconds to $30\einh{min}$ minutes due to the higher electrical resistance. The maximum current is below $25\einh{\upmu A/cm^2}$ including leak currents. 
As the bonds form over time, the sodium content in the glass decreases exponentially with a time constant of a few minutes. In \Figureref{current}(b), the currents from the first $8\einh{min}$ were scaled to get a smooth transition between the two regimes (dashed lines), which then fits an exponential decay. 
When the current arrives at an equilibrium, the bonding process has finished and only the remaining leak currents remain. 
Lastly the cell is cooled down slowly to avoid temperature gradients over the whole assembly.

The cell is filled by attaching the filling tube to a manifold and pumping it down to $10^{-7}\einh{mbar}$. A rubidium droplet is transferred into the cell and everything is heated for $6\einh{h}$ at $140\einh{^\circ C}$ and pumped for another $12\einh{h}$ to remove possible reaction products. After that the cell can be sealed by melting off the tube. The chromium electrodes are contacted from the outside with wires, to apply electric fields inside of the cell.

\begin{figure}[t]
\centerline{\includegraphics[width=8.4cm]{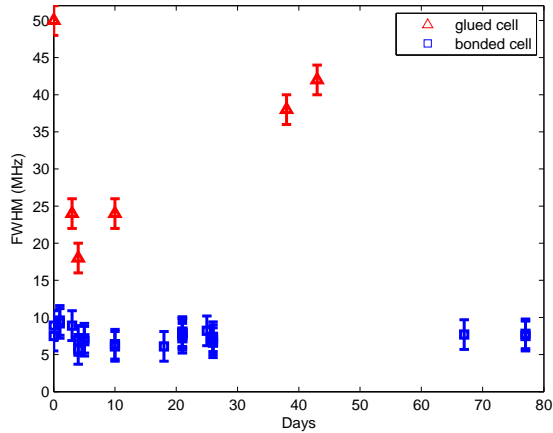}}
\caption{Linewidth in a glued cell (red triangles) and in a bonded cell (blue squares) versus time after filling the cell. In the glued cell a narrow linewidth was not possible due to outgassing of the glue. In the bonded cell a linewidth below $10\einh{MHz}$ is observed. An error of $2\einh{MHz}$ is assumed due to fit error, reference drift, and power fluctuations. \label{linewidth}
}
\end{figure}

\begin{figure}
\begin{overpic}[width=8.4cm, unit=1cm]{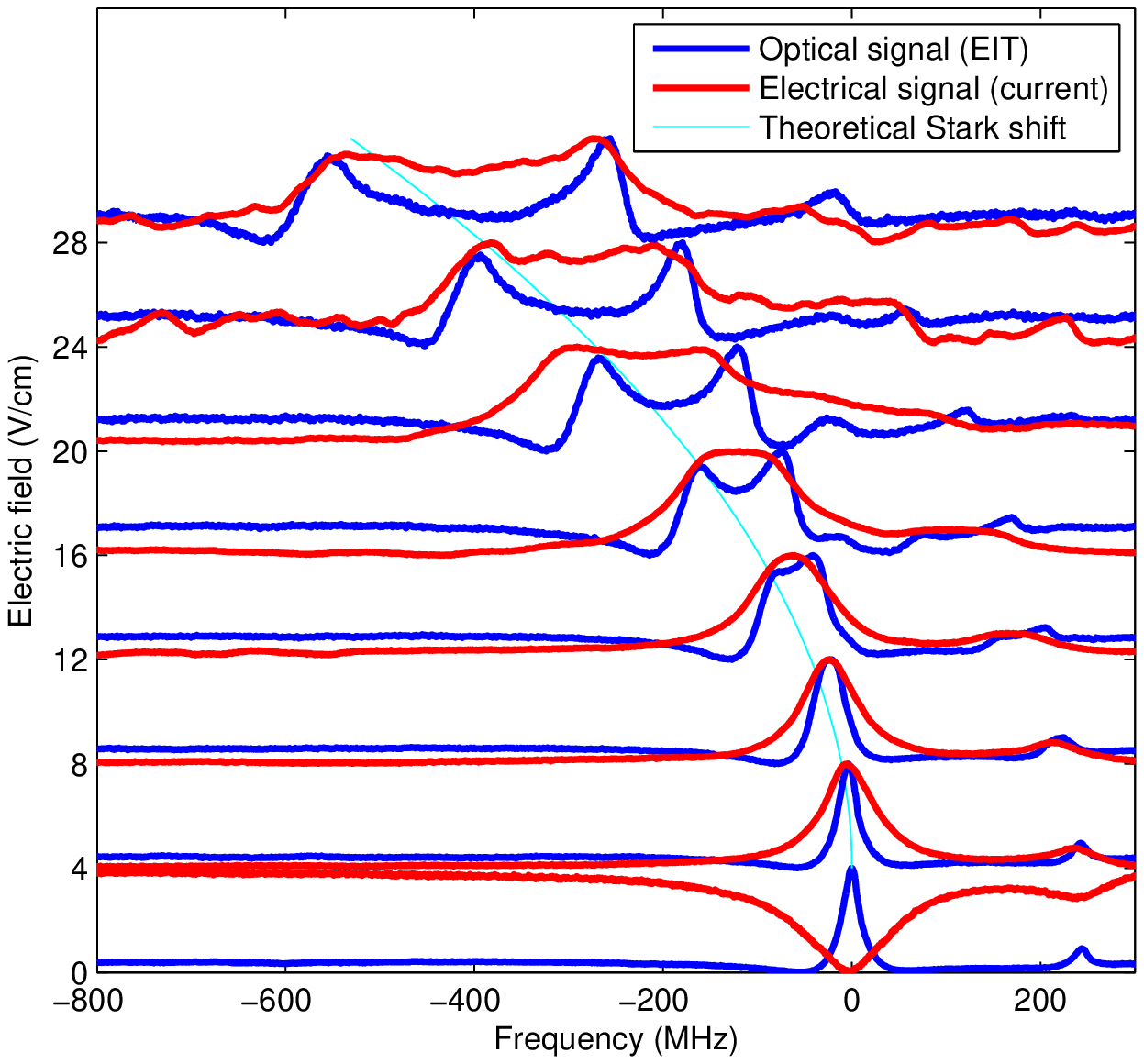}
\put(1.2,1.1){\includegraphics[width=3cm]{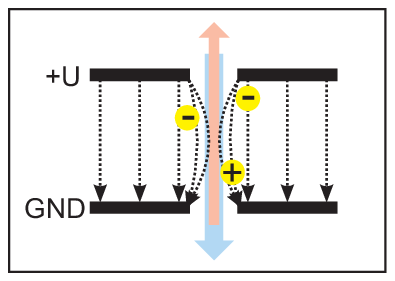}}
\end{overpic}
\caption{Normalized EIT signal and ionization current. DC Stark shift is observed in the bonded cells. The lineshapes are explained in the main text. The current signal for zero field is flipped because the direction of the current is determined by asymmetric charge production in the cell. The inset shows the electrode configuration, the two excitation laser beams and some charges inside the cell, which modify the electric field. \label{DC_Stark_shift}}
\end{figure}

The quality of the cells is tested by electromagnetically induced transparency (EIT) spectroscopy on the $5\mathrm{S}_{1/2}\rightarrow 5\mathrm{P}_{1/2}$ and the $5\mathrm{P}_{1/2}\rightarrow 30\mathrm{S}_{1/2}$ transitions. The two lasers are in a counter-propagating geometry perpendicular to the glass plates. As the chromium electrodes are not transparent the lasers are aligned between two neighboring electrodes, which are both connected to the same voltage (see inset of \Figureref{DC_Stark_shift}). As the distance between them is only one millimeter and the cell has a thickness of $5\einh{mm}$, the inhomogeneity created by this electrode configuration can be neglected.
Rydberg-EIT provides a very sensitive method to probe any residual background gas, as 
the background pressure inside the cell can be calculated from the pressure broadening of the narrow EIT peak.
In a glued reference cell we were able to achieve a minimum linewidth of $18\einh{MHz}$ and the lifetime of the cell was on the order of one month. By using the glue-free anodic bonding technique we are now able to achieve a linewidth of $6$ to $10\einh{MHz}$ depending on the rubidium vapor pressure.  The linewidth in the two types of cells is plotted over time in \Figureref{linewidth}. All these measurements were taken at temperatures between $60$ and $100\einh{^\circ C}$. After more than 4 months there was no measurable increase in background pressure. The cell was operated at maximum temperatures of $230\einh{^\circ C}$, only limited by our oven, without any increase of pressure broadening except from the vapor pressure of rubidium.

To test if the electrodes are still conductive and that the bonding did not destroy the feedthroughs, we applied DC voltages at the electrodes and a corresponding Stark shift of the Rydberg-EIT signal was observed. In addition to the optical measurement we performed Rydberg current measurements as in Ref. \onlinecite{bar2013}. Instead of probing the optical coherence the current is proportional to the excited Rydberg population. \Figureref{DC_Stark_shift} shows EIT (blue curve) and current (red curve) for the 30S state for different applied voltages in the bonded cell. 
Observing a Stark shift with a DC voltage clearly shows that the electrodes are still conductive. Otherwise charges would accumulate on the electrodes inside the cell and compensate the external voltage \cite{moh2007}. The fact that an ionization current from the atoms can be measured is even stronger proof of the good conductivity of the feedthroughs.

The lineshape of the signals can be explained taking space charges into account. They can be produced due to the ionization of the atoms or the photoeffect on the surfaces of the cell, e.g. if there is a thin rubidium layer. These charges create an additional inhomogeneous electric field, which shows a spatial dependence along the laser beams, shown in the inset of \Figureref{DC_Stark_shift}. For that reason the atoms  see different electric fields at different positions.  When we measure the Rydberg population using the ionization current, the signal from different positions and corresponding shifts is summed up to result in a flat top signal from minimum to maximum shift corresponding to minimum and maximum field present in the cell. In the case of EIT a peak can be compensated by the enhanced absorption of the neighboring peaks. In that case the signal can even cancel completely except for the signal from the minimum and the maximum field. 


In conclusion we fabricated a high vacuum tight vapor cell with thin film electrical feedthroughs where the pressure inside is only dominated by the vapor pressure of rubidium and not by background gases. We confirmed this low background pressure by a narrow Rydberg-EIT linewidth of $6\einh{MHz}$ and the quality of the thin film electrodes by DC Stark shift and ionization current measurements. The observed lineshape can be explained with an inhomogeneous electric field distribution inside the cell due to a high charge production.
The cell can withstand temperatures of at least $230\einh{^\circ C}$ and is stable for at least four months, while the spectroscopic results indicate a much larger lifetime. 


This technique can be combined with other cell fabrication techniques, like etched channels \cite{balu2010}. The complexity of the structures on the glass plates can be increased with state of the art LCD display fabrication techniques. Also, bonding silicon to the glass plates in a first step is possible and allows for SOI-based devices. Replacing the electrode material with graphene but keeping chromium for the feedthroughs might allow for highly transparent electrodes, which would still provide nicely homogenous electric fields.

We acknowledge the technical assistance of F. Schreiber and we thank the crystal preparation at the MPI Stuttgart. The work is supported by the ERC under contract number 267100 and BMBF within QuOReP (Project 01BQ1013).

%

%
%

\end{document}